\documentclass[prb,twocolumn,amsmath,amssymb]{revtex4}

\usepackage{graphicx}
\usepackage{dcolumn}
\usepackage{bm}

\begin{document}

\preprint{APS/123-QED}

\title{YbNiSi$_3$: a new antiferromagnetic Kondo lattice with strong exchange interaction}

\author{M. A. Avila}
\author{M. Sera}%
\author{T. Takabatake}%
\affiliation{%
Department of Quantum Matter, ADSM, Hiroshima University,
Higashi-Hiroshima 739-8530, Japan
}%

\date{\today}

\begin{abstract}
We report on the structural, thermodynamic and transport
properties of high-quality single crystals of YbNiSi$_3$ grown by
the flux method. This compound crystallizes in the SmNiGe$_3$
layered structure type of the $Cmmm$ space group. The general
physical behavior is that of a Kondo lattice showing an
antiferromagnetic ground state below $T_N=5.1$~K. This is among
the highest ordering temperatures for a Yb-based intermetallic,
indicating strong exchange interaction between the Yb ions, which
are close to +3 valency based on the effective moment of
$4.45$~$\mu_B/$f.u. The compound has moderately heavy-electron
behavior with Sommerfeld coefficient $190$~mJ/mol~K$^2$.
Resistivity is highly anisotropic and for I$\perp$b exhibits the
signature logarithmic increase below a local minimum, followed by
a sharp decrease in the coherent/magnetically ordered state,
resulting in residual resistivity of $1.5$~$\mu\Omega$~cm and
$RRR=40$. Fermi-liquid behavior consistent with a ground-state
doublet is clearly observed below 1 K.

\end{abstract}

\pacs{75.30.Mb,72.15.Qm,75.20.Hr,71.27.+a}

\keywords{ytterbium nickel silicides, mixed valency, heavy
fermions}

\maketitle

Ytterbium compounds continue to be subject of great interest due
to the variety of unusual physical properties they can present, in
general associated with the fact that this element's $f^{13}$ and
$f^{14}$ electronic states are very close in energy, and also
hybridize easily with the conduction band. As a consequence,
Yb-based compounds can display mixed valency, heavy fermion and
Kondo lattice characteristics, which provide opportunities for
better understanding the physics of such phenomena. In this sense,
Yb is often regarded as the ``hole'' equivalent of Ce whose
$f^{0}$ and $f^{1}$ states display the same characteristics, but
the latter's compounds are by far the most explored among these
two rare earths, in many cases simply because the Yb-based
compounds bear the 'stigma' of being considered more difficult to
synthesize\cite{fis92a}.

The RTX$_3$ family of intermetallics (R = rare earth; T =
transition metal; X = Si, Ge) is one such case. Many of the
Ce-based compounds in the family have been synthesized and
extensively investigated over the past decade or
so,\cite{hae85a,pec93a,yam95a,das97a,mur98a,kan99a,pik03a} but
practically nothing has been reported so far on heavy rare-earth
(R = Tb-Lu) members of these series\cite{gor77a,gla78a} and, in
particular, for R = Yb. We thus thought it would be worth
initiating such a line of research and here report on our first
results for the title compound.

\begin{figure}[tb]
\includegraphics[angle=0,width=88mm]{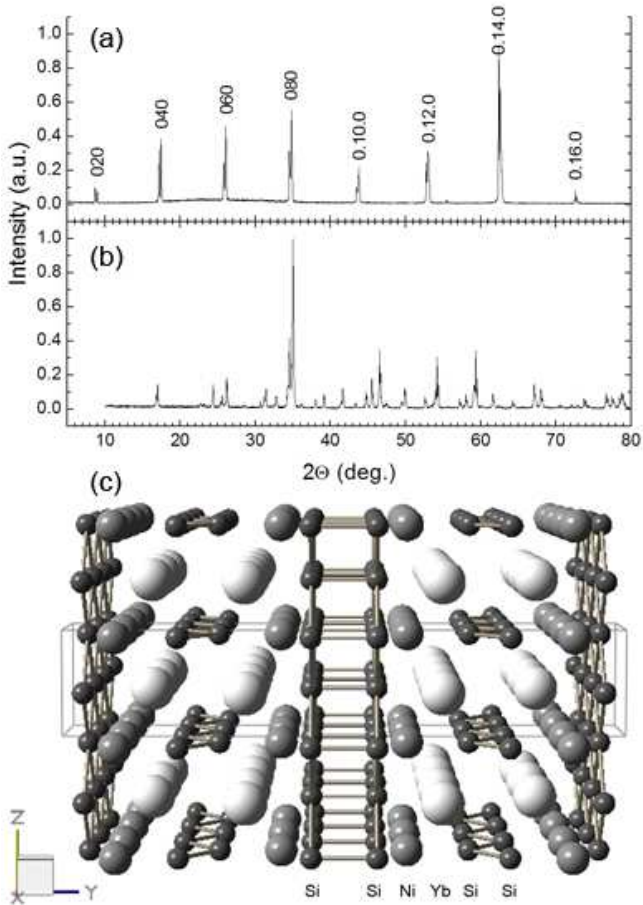}
\caption{\label{xray} (a) XRD (Cu K$\alpha$) pattern of a
plate-like crystal surface, revealing only the (0k0) reflections.
(b) XRD pattern of YbNiSi$_3$ crystals ground to a fine powder,
showing the full set of Bragg reflections. (c) Schematic
representation of the atomic positions showing the unit cell
(solid line) and layered structure along the $b$ axis.}
\end{figure}

Single crystals of YbNiSi$_3$ were grown from Sn flux using a
starting proportion of 1:1:3:20. The high-purity elements (Yb:
Ames Laboratory) were loaded and sealed inside an evacuated quartz
ampoule, which was then heated to 1150~$^\circ$C and slowly cooled
to 500~$^\circ$C, at which point the ampoule was removed from the
furnace and most of the excess flux was separated by decanting.
The crystals have excellent resistance to acid and can be left in
a pure HCl bath for as long as necessary to remove any remaining
flux droplets from the surface. The resulting crystals are
plate-like, with the main surface orthogonal to the
crystallographic $b$ direction, as evidenced from the surface
x-ray diffraction (XRD) pattern shown in Fig.~\ref{xray}(a) which
has only (0k0) reflections. The largest crystal plates were
limited mostly by the ampoule wall itself, but the best crystals -
those that grew compact, isolated, with smooth surfaces and
straight edges - have lengths up to 5~mm and thickness up to
0.3~mm. Many of the crystals display some interesting, rectangular
surface topology patterns resulting from the particular
flux-growth dynamics of this compound. Electron-probe
microanalysis (EPMA) confirmed the stoichiometric 1:1:3 proportion
in the crystals and found the Sn inclusions to be less than
0.03\%. Our attempts to grow non-magnetic YNiSi$_3$ and LuNiSi$_3$
crystals by the same method were unsuccessful, so the phase
diagram seems to be particularly favorable for R = Yb.

Fig.~\ref{xray}(b) shows the XRD pattern of crystals that were
crushed into a fine powder. The pattern is consistent with the
$Cmmm$ space group of the SmNiGe$_3$ structure type shown in
Fig.~\ref{xray}(c), which has close values for the $a$ and $c$
unit cell parameters and a layered structure along the $b$ axis. A
Rietveld refinement of this spectrum resulted in
$a=3.8915(1)$~\AA, $b=20.8570(6)$~\AA, $c=3.9004(1)$~\AA, and
$V=316.58$~\AA$^3$. The crystals' plate-like morphology described
above is therefore consistent with the general observation that
the macroscopic dimensions of flux-grown crystals often have an
inverse relation to the microscopic lattice parameters. A more
detailed structural study will be required though, in order to
determine the exact atomic positions in the unit cell of
YbNiSi$_3$.

\begin{figure}[htb]
\includegraphics[angle=0,width=88mm]{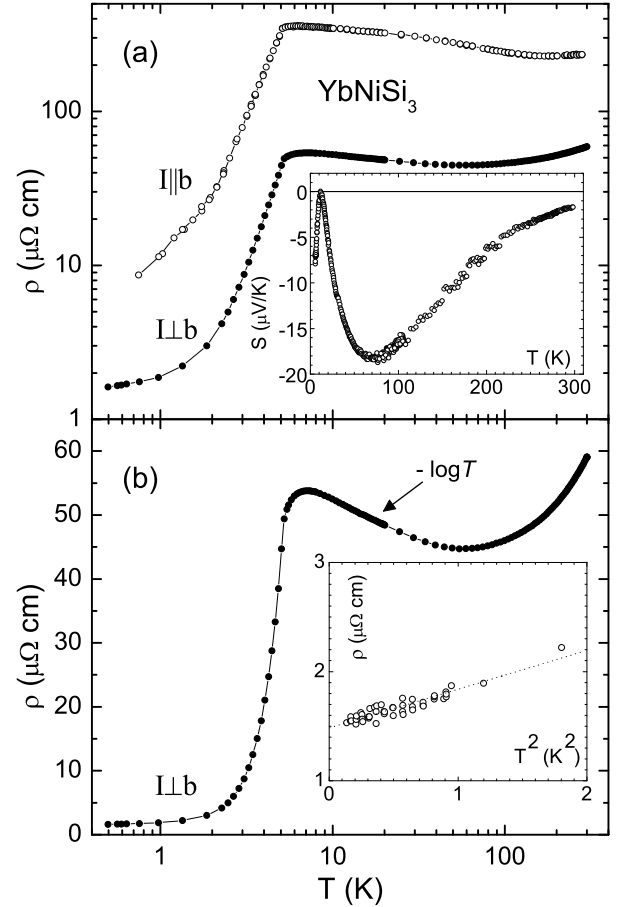}
\caption{\label{transp} (a) Anisotropic electrical resistivity
$\rho(T)$ of YbNiSi$_3$ below room temperature. The inset shows
the thermopower $S(T)$. (b) Log$T$ dependence of $\rho(T)$ for
I$\parallel$b. The inset shows the $T^2$ dependence of $\rho(T)$
at temperatures below 1~K.}
\end{figure}

As one might expect from the layered structure of YbNiSi$_3$, the
electrical resistivity below room temperature is anisotropic as
shown in Fig.~\ref{transp}(a). For I$\perp$b the behavior is
metallic, with room temperature resistivity of about
60~$\mu\Omega$~cm which initially decreases on cooling, reaches a
local minimum centered at 55~K, then increases again until
reaching a peak at 7~K. Below this temperature it begins to
decrease very fast. A peak in $d\rho/dT$ places the maximum slope
of this drop at 5.0~K, which we will later show to be associated
with the Ne\'{e}l temperature $T_N$ of the antiferromagnetically
ordered Yb moments, and therefore the strong decrease as a whole
can be understood as a combined effect of magnetic ordering at
$T_N$ with the onset of coherent scattering of the hybridized Yb
moments below $T\sim7$~K. Figure~\ref{transp}(b) shows the
I$\perp$b resistivity plotted against log$T$, where a logarithmic
behavior is seen between the local minimum and the peak, a
signature of the Kondo effect. The inset of Fig.~\ref{transp}(b)
details the resistivity behavior well below the transition,
plotted against $T^2$. A Fermi-liquid type $\rho=\rho_0+AT^2$
behavior is clearly observed below 1~K, with
$\rho_0=1.48$~$\mu\Omega$~cm and $A=0.36$~$\mu\Omega$~cm/K$^2$.
The residual resistivity ratio ($RRR$) defined as
$\rho(300K)/\rho_0$ is 40. These results attest the clean metallic
character and high crystallographic quality of the sample.

The resistivity measurements with I$\parallel$b, shown in
Fig.~\ref{transp}(a), were made rather difficult due to the fact
that the crystals do not grow large in this direction and we were
only able to place 4 contacts on a crystal cut to form a rather
irregularly shaped bar along the $b$ axis. Therefore the estimate
of the sample cross-section may contain errors of a factor of 2 or
even more. Still, it is likely that in this direction the
resistivity level is significantly higher than for I$\perp$b and
the temperature dependence resembles that of a semi-metal,
initially almost flat and then increasing down to the transition
temperature. This change in qualitative behavior may be a
manifestation of a strongly anisotropic Fermi surface resultant
from the layered structure, since in this direction the conduction
band crosses tightly bound Si double-layers as shown in
Fig.~\ref{xray}(c).

The inset of Fig.~\ref{transp}(a) shows the measured thermopower
$S(T)$. As for most Yb compounds displaying mixed valent behavior,
$S(T)$ shows a broad minimum below room temperature. The minimum
is located at 72~K and reaches a value of $S=-19~\mu$V/K. Below
this temperature it increases again, reaches a peak very close to
$S=0$ at 12~K and from there $S(T)$ starts decreasing fast again.
Since it is expected to vanish at $T=0$ we can deduce that a
second minimum exists. The peak at 12~K places YbNiSi$_3$ in the
crossover region between compounds where Yb is essentially +3 and
such a peak enters the positive side of $S(T)$, and compounds
where Yb displays intermediate valency behavior and this peak is
either well into the negative region or simply doesn't
exist.\cite{jac82a,yad99a}

\begin{figure}[htb]
\includegraphics[angle=0,width=88mm]{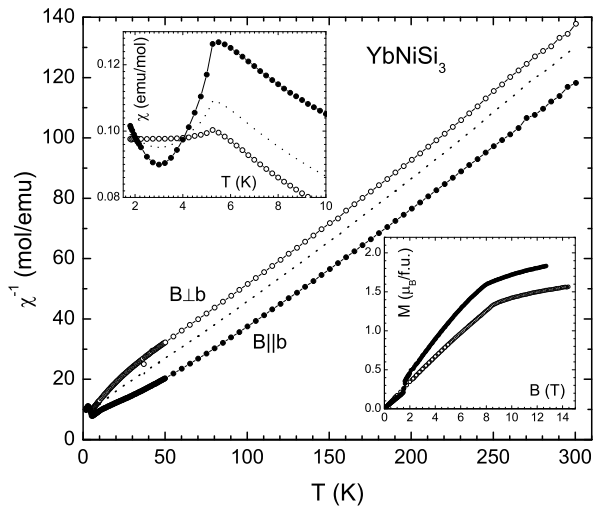}
\caption{\label{mag} Anisotropic magnetic behavior of YbNiSi$_3$.
Solid symbols are for B$\parallel$b, open symbols are for
B$\perp$b, and dotted lines are the polycrystalline averages. Main
plot: temperature dependence of the inverse susceptibility at
$B=0.1$~T, showing Curie-Weiss behavior at high temperatures.
Upper inset: detail of the antiferromagnetic transition at
$T_N=5.1$~K appearing in susceptibility. Lower inset:
magnetization isotherms at $T=2$~K, revealing a spin-flop
transition at 1.7~T for $B\parallel$b.}
\end{figure}

The magnetic behavior is also anisotropic since the CEF
environment at the Yb site splits the $4f$ multiplet into
non-degenerate electronic energy levels. Well above the magnetic
ordering temperature, the inverse susceptibility $\chi^{-1}(T)$
for $B=0.1$~T is typical of paramagnetic local moments with CEF
effects (Fig.~\ref{mag}). The $\chi^{-1}(T)$ curves for
\textbf{B}$\parallel$b and \textbf{B}$\perp$b are essentially
linear and parallel to each other, and the easy alignment
direction for the Yb moments is towards the $b$ direction. As the
temperature decreases below 50~K the two curves deviate strongly
from Curie-Weiss behavior, while the polycrystalline average
estimated as $\chi^{-1}_p=3/(\chi_b+2\chi_{ac})$ (dotted line)
which should average out the CEF effects assuming that any
existing anisotropy within the $ac$ plane is small, continues to
be essentially linear almost down to the ordering temperature. A
fit of the polycrystalline average curve above 150~K to the
Curie-Weiss law results in an effective moment
$\mu_{eff}=4.45$~$\mu_B/$f.u. (very close to the expected value of
a Yb$^{+3}$ moment) and $\Theta_p=-11.6$~K (indicative of
antiferromagnetic coupling).

The upper inset in Fig.~\ref{mag} details the anisotropic
antiferromagnetic transition in the susceptibility $\chi(T)$ at
$B=0.1$~T. A peak in $d(\chi T)/dT$ gives $T_N=5.0$~K for this
field. The behavior of both curves in the magnetically ordered
state is indicative of a non-trivial, and possibly canted,
arrangement of the magnetic moments. The upturn at low
temperatures for \textbf{B}$\parallel$b appears to be a genuine
response of the compound and not a trivial impurity effect, first
because it does not appear in the \textbf{B}$\perp$b curve, and
second because it was reproduced almost exactly in two different
batches of samples grown with different elemental purities (3N in
the first batch and 4N-5N in the second.)

The lower inset in Fig.~\ref{mag} shows magnetic isotherms at 2~K
for both directions up to 14~T. There is a spin-flop transition at
$B=1.7$~T in the \textbf{B}$\parallel$b curve. A change in slope
near 8~T of both curves marks the crossing of the $T_N(B,T)$ line
(a detailed phase diagram will be presented in a future
communication), but the magnetization still maintains a positive
slope above this transition. If no other meta-magnetic transitions
are to be found at even higher fields, then it is clear from the
graph that the saturated moment at 2~K should remain below
2~$\mu_B/$f.u. and therefore significantly smaller than the
expected value of 4~$\mu_B/$f.u. for saturated Yb$^{+3}$ moments.

\begin{figure}[htb]
\includegraphics[angle=0,width=88mm]{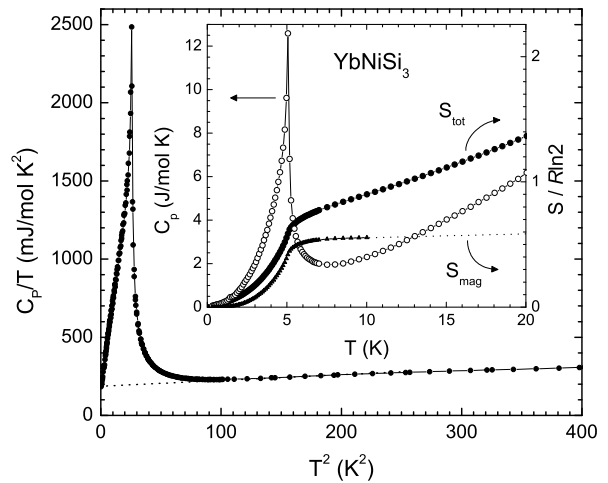}
\caption{\label{heat} Zero-field heat capacity measurement on
YbNiSi$_3$ at low temperatures, presented as $C_p/T~vs.~ T^2$. The
dotted line is a linear extrapolation from the 10-20~K region. The
inset shows the original $C_p~vs.~ T$ data ($\circ$), the total
entropy $S_{tot}$ ($\bullet$) and the magnetic entropy $S_{mag}$
($\blacktriangle$).}
\end{figure}

In order to shed further light on the ground state characteristics
of YbNiSi$_3$, the low temperature, zero-field heat capacity was
measured on a Quantum Design PPMS system with $^3$He option. The
open symbols in the inset of Fig.~\ref{heat} show the measured
$C_p(T)$ data below 20~K. A very sharp lambda-like peak is seen at
$T_N=5.1$~K.

The main graph of Fig.~\ref{heat} shows the same data presented as
$C_p/T~vs.~ T^2$. A linear region is observed between 20 and 10~K,
and its extrapolation to $T=0$ (dotted line) coincides with the
levelled value of $\gamma=190$~mJ/mol~K$^2$ that is reached by the
lowest temperature data below 1~K. The coincidence of these two
independent estimations gives us confidence in claiming that the
obtained value is a good representation of the electronic specific
heat, therefore placing it as a moderately heavy-electron system.
From the slope of the linear region we obtain the lattice
coefficient $\beta=0.26$~mJ/mol~K$^4$ and estimate the Debye
temperature as $\Theta_D=330$~K. We can infer that the first
excited CEF levels' energy scale should lie well above 20~K, since
no evidence of Schottky anomalies are seen up to this temperature.

In the inset we also present the total entropy $S_{tot}(T)$ (solid
symbols) obtained by numerical integration of $C_p/T~vs.~ T$ and
an evaluation of the magnetic entropy $S_{mag}(T)$, which was
obtained by removing the electronic and lattice contributions
estimated above from the specific heat before integration. In
either case it becomes clear that the total entropy accumulated up
to $T_N$ is $~0.6R$ln2. Thus, we may conclude that a ground state
doublet is responsible for the magnetic ordering, and the Yb
moments are already significantly screened when the magnetic
ordering ensues. With the obtained value of $\gamma$ we can apply
Rajan's expression\cite{raj83a} for the specific heat of the
Coqblin-Schrieffer model $\gamma T_K\approx11.2j$ to estimate the
Kondo temperature of YbNiSi$_3$ ($j=1/2$) as $T_K=30~K$.

Finally, we can also use $\gamma$, in combination with the
previously calculated $A$ coefficient of the Fermi-liquid
resistivity model, to estimate
$A/\gamma^2=10^{-5}~\Omega$~cm~(mol~K/J)$^2$, which is the
well-known Kadowaki-Woods ratio, quite commonly observed in heavy
fermion systems featuring a ground-state
doublet.\cite{tsu03a,kon03a}

In many cases where magnetic ordering coexists with mixed valent,
Kondo lattice or heavy fermion behavior, quantitative analysis of
the material's low temperature properties becomes rather difficult
due to the convolution of several contributions to the
thermodynamic and transport properties, a problem which can be
further aggravated by difficulties in preparation of samples of
high quality. The results we have obtained demonstrate that
YbNiSi$_3$ is a welcome exception to the rule and several physical
parameters could be quantitatively determined with good accuracy
from our measurements. Furthermore, this compound was found to
belong to the very rare group of Yb-based materials with magnetic
ordering above 5~K, together with YbPtAl\cite{die95a},
YbNiSn\cite{dre96a} and YbB$_2$,\cite{avi03a} indicating that the
intersite indirect magnetic exchange interaction between local
moments is strong and dominant, while still displaying very
characteristic features of a system with strong on-site Kondo
interaction. Therefore, we believe that YbNiSi$_3$ presents itself
as a promising model system for more in-depth investigations on
the delicate balance between these two competing energies, as well
as other issues related to Yb compounds and Kondo lattices in
general. Studies on the high-field properties and magnetic phase
diagram, as well as physical and chemical pressure effects in this
compound are currently in progress.\\
\\

\begin{acknowledgments}
We are thankful to Y. Shibata for the EPMA analysis, to T.
Sasakawa for assistance with the XRD refinements, and to F. Iga
for his help in the specific heat measurements. The
low-temperature measurements were performed at the Materials
Science Center, N-BARD, Hiroshima University. This work was
supported by a Grant-in-Aid for Scientific Research (COE Research
13CE2002) of MEXT Japan.
\end{acknowledgments}


\end{document}